# Single-photon spin-orbit entanglement violating a Bell-like inequality

**Lixiang Chen and Weilong She***

*State Key Laboratory of Optoelectronic Materials and Technologies, Sun Yat-sen University, Guangzhou 510275, China*

*\*Corresponding author: shewl@mail.sysu.edu.cn*

Single photons emerging from q-plates (or Pancharatnam-Berry phase optical element) exhibit entanglement in the degrees of freedom of spin and orbital angular momentum. We put forward an experimental scheme for probing the spin-orbit correlations of single photons. It is found that the Clauser-Horne-Shimony-Holt (CHSH) parameter *S* for the single-photon spin-orbit entangled state could be up to $2\sqrt{2}$, evidently violating the Bell-like inequality and thus invalidating the noncontextual hidden variable (NCHV) theories.

*OCIS codes: 230.5440, 270.5585*

## 1. Introduction

Besides spin angular momentum, single photons can carry another new degree of freedom, namely, orbital angular momentum (OAM) [1]. The OAM carried by a twisted photon is associated with the helical wave front, which opens the possibility for encoding a quNit with a single photon [2, 3]. Traditionally, it was regarded that there is no interaction between spin and OAM. However, the translation between spin and OAM was recently found to be possible and raised much research interest [4-10]. For example, it was demonstrated that the conversion from spin to orbital angular momentum could be realized by using q-plates (or Pancharatnam-Berry phase optical elements) [4-8]. Besides, quantum controlled NOT gates with spin and OAM of single-photon quantum logic has also been demonstrated [11]. On the other hand, it has been recognized that using several degrees of freedom of a single-photon to encode multiple qubits had potential in building a deterministic quantum information processor [12-15]. Actually, the entanglement is not limited to different particles and is generally applicable to different degrees of freedom in single particles [16,17]. The scheme for testing the validity of noncontexual hidden variable (NCHV) with single-particle two-qubit states has been proposed [18]. Further experiment with single-photon entanglement in polarization and momentum [19,20] or with single-neutron entanglement in spin and path [21] has shown the violation of a Bell-like equality and therefore invalidated the NCHV theories. Here, we put forward another experimental scheme for testing the violation of a Bell-like inequality with single-photon spin-orbit entangled state that is created by a q-plate.

## 2. The q-plates

According to quantum mechanics, the single-photon spin-orbit product states can be described by a tensor product Hilbert space, namely $H = H_1 \otimes H_2$, where $H_1$ and $H_2$ are two disjoint Hilbert spaces corresponding to spin and OAM degrees of freedom, respectively. To produce the spin-orbit entanglement, we should realize the coupling between two vector-states lying in spin and OAM Hilbert spaces. The natural solution emerges from an old physical principle: the so-called Pancharatnam-Berry optical phase [22,23]. One important feature of the Pancharatnam-Berry phase is that the input polarization (spin) can reshape the output wavefront (orbit).

Recently, a novel device called q-plate was built, which can reverse the spin of photons while transferring the change of spin angular momentum into the orbital kind [4]. A general q-plate is a planar slab of a uniaxial birefringent medium, with an inhomogeneous orientation of the optical axis lying in x-y plane and a homogeneous phase retardation of $\pi$ along z-axis. The orientation of optical axis in a polar coordinate can be described by $\alpha(r,\varphi) = q\varphi + \alpha_0$, where $q$ and $\alpha_0$ are two constant. Some types of q-plates are illustrated in Fig. 1 and one can see that the q-plate with parameter $q$ has a $|2(q-1)|$-fold rotational symmetry [24]. Based on q-plates, increase of Shannon dimensionality [25], entanglement transfer from spin to OAM [26], and quantum cloning of OAM qubits [27] were recently reported. On the circular bases, the transmitted matrix for the q-plate can be expressed as

$$T_q = \begin{bmatrix} 0 & \exp[i2\alpha(r,\varphi)] \\ \exp[-i2\alpha(r,\varphi)] & 0 \end{bmatrix}. \quad (1)$$

In a general case, the incoming photon is of spin and OAM state, namely $|\psi\rangle = (\alpha|L\rangle + \beta|R\rangle) \otimes |m\rangle$, where $|L\rangle$ and $|R\rangle$ are left- and right-handed circular polarizations with two constants $\alpha$ and $\beta$ ($|\alpha|^2 + |\beta|^2 = 1$) describing an arbitrary polarization state; and $|m\rangle$ denote the eigenstates of the OAM operator. Thus, in the single photon space the q-plate can be described by a quantum operator as follows

$$\hat{Q}(q) = \exp(i2\alpha_0)|R, m+2q\rangle\langle L, m| + \exp(-i2\alpha_0)|L, m-2q\rangle\langle R, m|. \quad (2)$$

Assume that the input photons are horizontal linearly polarized $|H\rangle = (|L\rangle + |R\rangle)/\sqrt{2}$, and then the output from the q-plate will be a spin-orbit Bell state $|\psi\rangle = (|R, +2q\rangle + |L, -2q\rangle)/\sqrt{2}$. As was pointed out, this state exhibits single-photon spin-orbit entanglement [28]. Interestingly, such a spin-controlled OAM generation can also be visually understood as a correspondence between the points on the spin Poincaré-sphere and those on the orbital Poincaré-sphere of $|2q|$ modes [29]. For $q > 0$, the spin and $|2q|$-oder orbit Poincaré-spheres overlap each other, while for $q < 0$, the two Poincaré-spheres are of inversion symmetry with each other. In the following, we will demonstrate an experimental scheme to probe the strong spin-orbit correlation in single photons, and predict the violation of a Bell-like inequality.

## 3. Single-photon spin-orbit entanglement violating Bell-like inequality

Local theories represent a particular circumstance of non-contextuality, in that the result is assumed not to depend on measurements made simultaneously on spatially separated (mutually non-interacting) systems. In order to test non-contextuality, joint measurements of commuting observables that are not necessarily separated in space are required. Generally, the experiment for testing single-photon entanglement consists of three stages: preparation, manipulation, and detection of the entangled state. Our experimental scheme was sketched in Fig. 2, which is inspired by Michler *et al* [19]. We employ the type-I spontaneous parametric down-conversion (SPDC) as the Einstein-Podolsky-Rosen (EPR) source that produces twin photons in polarization entangled state $|\Psi\rangle_{spin} = (|H\rangle_A |H\rangle_B + |V\rangle_A |V\rangle_B)/\sqrt{2}$. The single-mode fiber (SMF) in Bob's side exclusively sustains the fundamental mode with zero OAM ($m = 0$) and thus cancels out the possible OAM entanglement [30]. As the eigenmodes in q-plates are circular polarizations, after substituting the relation $|H\rangle = (|L\rangle + |R\rangle)/\sqrt{2}$ and $|V\rangle = (|L\rangle - |R\rangle)/i\sqrt{2}$, we can rewrite the polarization entangled state as $|\Psi\rangle_{spin} = (|L\rangle_A |R\rangle_B + |R\rangle_A |L\rangle_B)/\sqrt{2}$. From Eq. (2), we know that this state, after passing through the q-plate with $q=1$ [see Fig. 1(a)], will show a hybrid pattern:

$$|\Psi\rangle_{Hybird} = \frac{1}{\sqrt{2}}[|L\rangle_A (|L\rangle \otimes |m=-2\rangle)_B + |R\rangle_A (|R\rangle \otimes |m=+2\rangle)_B]. \tag{3}$$

If we measure the photons at Alice's hand in the basis $|H\rangle$, we know that Bob's photons will simultaneously collapse into the single-photon spin-orbit entangled state

$$|\psi\rangle_B = \frac{1}{\sqrt{2}}(|L\rangle \otimes |m=-2\rangle + |R\rangle \otimes |m=+2\rangle)_B. \tag{4}$$

So the measurement at Alice's hand can serve as the trigger signal and the coincidence event therefore assures that Bob does work at a single photon level. Up to now we have successfully prepared the single-photon spin-orbit entangled state. Similarly to [19,20], to verify the existence of single-photon spin-orbit entanglement, the expectation value $E(\chi_A, \chi_B)$ of the joint measurement for the OAM state $[|m=-2\rangle \pm \exp(i\chi_A)|m=+2\rangle]/\sqrt{2}$ and the spin state $[|H\rangle \pm \exp(i\chi_B)|V\rangle]/\sqrt{2}$ should be manipulated by adjusting two parameters $\chi_A$ and $\chi_B$. However, it is worth noting that in our case it is spin-orbit entanglement rather than polarization-momentum or spin-path entanglement as reported in [19-21]. In the next, we will employ the Dove prisms to perform the direct OAM manipulation, and therefore excluding the polarization-momentum or spin-path entanglement.

Fig. 2

To manipulate the spin and OAM states by conventional polarizing beams splitter (PBS) rather than circular polarization beam splitter, we use the 45° orientated quarter-wave plate (QWP@45) to perform the transforms: $|L\rangle \rightarrow (1+i)/\sqrt{2}|H\rangle$ and $|R\rangle \rightarrow (1-i)/\sqrt{2}|V\rangle$ before sending Bob's photons into the interference stage. The first PBS@0 is set so that it transmits the polarization $|H\rangle$ but reflects polarization $|V\rangle$. Two Dove prisms embedded respectively in two arms are set so that they have a relative rotation angle of $\alpha$, and thus impart these two components $|H\rangle$ and $|V\rangle$ with a phase difference of $\chi_A = 2m\alpha$ (here $m = 2$) while not affecting their polarization states [31]. Then the non-polarizing beam splitter (BS) performs the projection onto two states $|\pm 1, \varphi_A\rangle = (1/\sqrt{2})[(1+i)|m=-2\rangle \pm (1-i)\exp(i\chi_A)|m=+2\rangle]$. Of special importance here is the employment of Dove prisms, which assures that the effect concerned is purely arising from phase difference induced by OAM rather than due to the path difference, say, for non-twisted photons with zero OAM, no phase difference between two arms is introduced when rotating one of Dove prisms. Our scheme evidently excludes single-particle polarization-momentum entanglement or spin-path entanglement [19-21]. The eigenvalues $\pm 1$ of the dichotomic observable, $\hat{A}(\chi_A) = |+1, \chi_A\rangle\langle +1, \chi_A| - |-1, \chi_A\rangle\langle -1, \chi_A|$, correspond to the detection of photon appearing in the ports $A^+$ and $A^-$, respectively. Placed in each output port of BS, the $-45°$ orientated quarter-wave plate (QWP@-45), as the complement of the aforementioned QWP@45, restores the left- and right-handed circular polarizations from the linear ones. Subsequently, two half-wave plates (HWP1@0 and HWP2@β) bring a phase difference $\chi_B = 2\beta$ between two components $|L\rangle$ and $|R\rangle$. Finally, the PBS2@0 or PBS3@0 performs the projection onto two states $|\pm 1, \chi_B\rangle = (1/\sqrt{2})[|L\rangle \pm \exp(i\chi_B)|R\rangle]$. Similarly, the observable $\hat{B}(\chi_B)$ is defined as $\hat{B}(\chi_B) = |+1, \chi_B\rangle\langle+1, \chi_B| - |-1, \chi_B\rangle\langle-1, \chi_B|$, following the pattern of $\hat{A}(\chi_A)$. Now we can calculate the expectation values $E(\chi_A, \chi_B)$ as

$$E(\chi_A, \chi_B) = {}_B\langle\psi|\hat{A}(\chi_A)\hat{B}(\chi_B)|\psi\rangle_B = \sin(\chi_A + \chi_B)$$
$$= \frac{N(A^+B^+) + N(A^-B^-) - N(A^+B^-) - N(A^-B^+)}{N(A^+B^+) + N(A^-B^-) + N(A^+B^-) + N(A^-B^+)}, \quad (5)$$

where $N(A^iB^j)$ $(i, j = \pm)$ denotes single-photon counts recorded by a detector in port $A^iB^j$.

Fig. 3

The concept of quantum non-contextuality represents a straightforward extension of the classical view: the result of a particular measurement is determined independently of previous (or simultaneous) measurements on any set of mutually commuting observables [32]. In order to test the non-contextuality, the spin and OAM degrees of freedom in single photons are prepared in a non-factorized state and are manipulated to obtain the joint measurements of two commuting observables, as shown in Eq. (5). Generally, the Bell-like inequality is introduced to test non-contextuality [18-21], which is given by $S = E(\chi_A, \chi_B) + E(\chi_A, \chi_B') - E(\chi_A', \chi_B) + E(\chi_A', \chi_B')$ with $-2 \leq S \leq 2$ [33]. However, quantum theory predicts that $S$ could be larger than 2. In an actual implementation, we inspect the single-photon counts $N(A^iB^j)$ by varying phase $\chi_A$ (i.e., by rotating the Dove prisms by a relative angle $\alpha = \chi_A/2m$) for two fixed phases $\chi_B$. The Bell-like inequality is valued based on $N(A^iB^j)$. By choosing $\chi_A = \pi/2$ ($\alpha = 22.5°$), $\chi_A' = -\pi$ ($\alpha = -45°$), $\chi_B = \pi/4$ (i.e., $\beta = 22.5°$), and $\chi_B' = -\pi/4$ ($\beta = -22.5°$), as illustrated in Fig. 3, one will find that the parameter $S$ can be up to $2\sqrt{2}$, evidently violating the Bell-like inequality, and therefore, invalidating the NCHV theories.

## 4. Conclusion

We have put forward an experimental scheme, in which single-photon spin-orbit Bell state is prepared and manipulated. The violation of a Bell-like inequality ($S = 2\sqrt{2}$) is predicted, which appears to show the invalidation of the NCHV theories.

## Acknowledgements

L Chen appreciates helpful discussions with Prof. L Marrucci and thanks Prof. S J van Enk very much for his kind help. The authors gratefully acknowledge the financial support from the National Natural Science Foundation of China (Grant No. 10874251).## References

1. L. Allen, M. W. Beijersbergen, R. J. C. Spreeuw, and J. P. Woerdman, "Orbital angular momentum of light and the transformation of Laguerre-Gaussian laser modes," Phys. Rev. A **45**, 8185-8189 (1992).
2. S. Franke-Arnold, L. Allen, and M. Padgett, "Advances in optical angular momentum," Laser & Photon. Rev. **2**, 299-313 (2008).
3. G. Molina-Terriza, J. P. Torres, and L. Torner, "Twisted photons," Nature Phys. **3**, 305-310 (2007).
4. L. Marrucci, C. Manzo, and D. Paparo, "Optical spin-to-orbital angular momentum conversion in inhomogeneous anisotropic media," Phys. Rev. Lett. **96**, 163905 (2006).
5. G. F. Calvo and A. Picon, "Spin-induced angular momentum swithching," Opt. Lett. **32**, 838-840 (2007).
6. E. Karimi, B. Piccirillo, L. Marrucci, and E. Santamato, "Light propagation in a birefringent plate with topological charge," Opt. Lett. 34, 1225-1227 (2009)
8

**Figure Captions**

Fig. 1. (Color online) Illustration of some types of $q$-plates, in which the orientation of optical axis is locally tangent to the line: (a) is rotational invariance; while (b), (c), (d), and (e) are of one-, two-, three-, and four-fold rotational symmetry, respectively.

Fig. 2. (Color online) Proposed experimental setup for testing the violation of the Bell-like SHCH inequality for single-photon spin-orbit entanglement: SMF, single-mode fiber; QWP, quarter-wave plate; PBS, polarizing beam splitter; M, mirror; DP, Dove prism; HWP, half-wave plate; BS, non-polarizing beam splitter; D, single-photon detector.

Fig. 3. (Color online) The theoretical single-photon counts $N(A^i B^j)$ when varying the phase $\chi_A$ while $\chi_B$ is fixed at $-\pi/4$ or $\pi/4$. The four circles denote the cases, at which $S$ could be up to $2\sqrt{2}$.

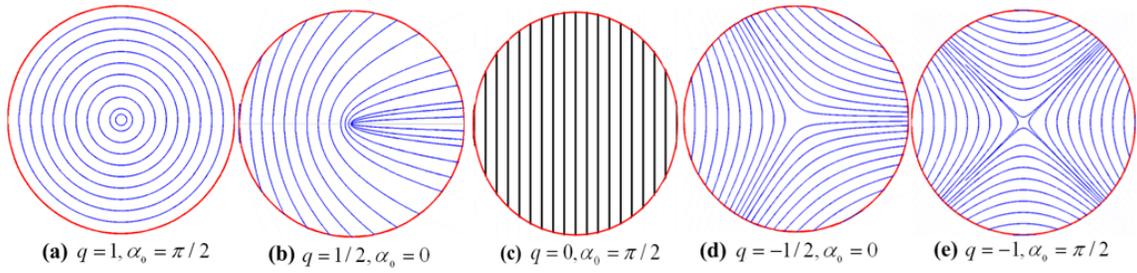

Fig. 1 (CHEN)

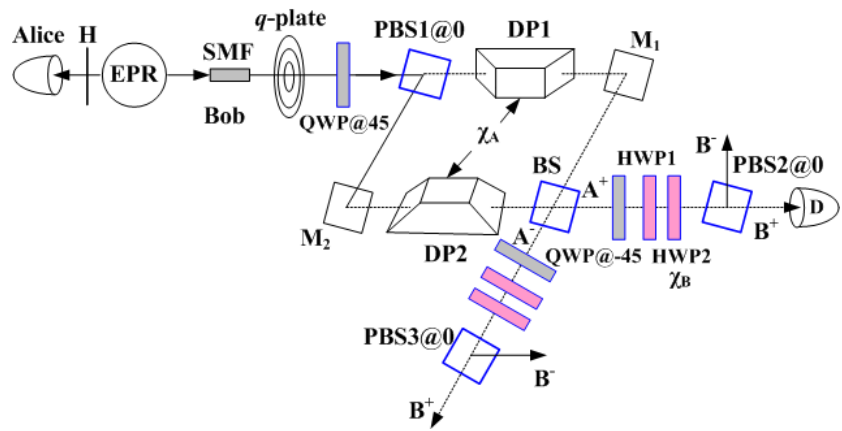

Fig. 2 (CHEN)

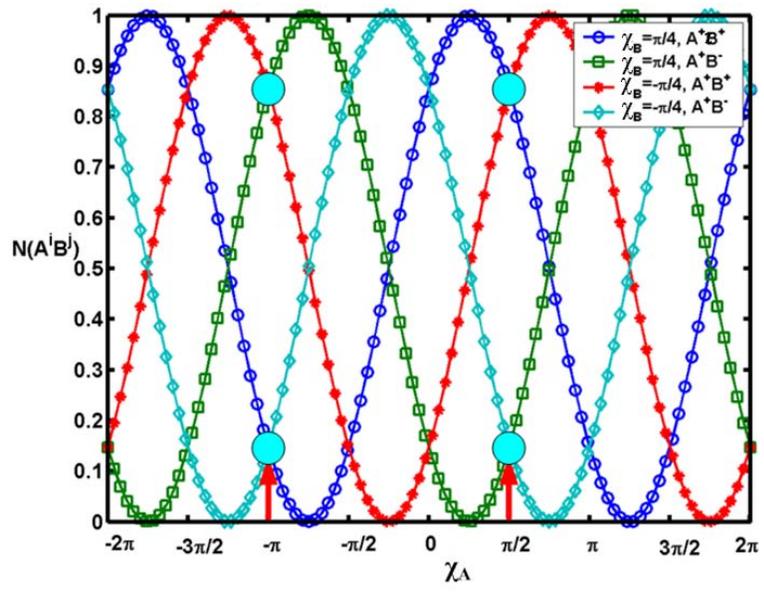

Fig. 3 (CHEN)